\documentclass[10pt, conference, compsocconf]{IEEEtran}

\ifCLASSINFOpdf

\else
 
\fi

\usepackage{cite}
\usepackage[cmex10]{amsmath}
\usepackage{caption}
\usepackage{tikz}
\usetikzlibrary{fit,positioning}
\usepackage{algorithm}
\usepackage{float}
\usepackage{algpseudocode}
\usepackage{booktabs}

\begin{document}
\title{Predicting diverse M-best protein contact maps}
\author{\IEEEauthorblockN{Siqi Sun,
Jianzhu Ma, Sheng Wang and
Jinbo Xu}
\IEEEauthorblockA{Toyota Technological Institute at Chicago,\\ Chicago,
IL, USA, 60637\\
Email: siqi.sun, majianzhu, wangsheng@ttic.edu,
jinbo.xu@gmail.com}}

\maketitle

\maketitle

\begin{abstract}
%\boldmath
\noindent Protein contacts contain important information for protein structure and functional study, but contact prediction from sequence information remains very challenging. Recently evolutionary coupling (EC) analysis, which predicts contacts by detecting co-evolved residues (or columns) in a multiple sequence alignment (MSA), has made good progress due to better statistical assessment techniques and high-throughput sequencing. Existing EC analysis methods predict only a single contact map for a given protein, which may have low accuracy especially when the protein under prediction does not have a large number of sequence homologs. Analogous to ab initio folding that usually predicts a few possible 3D models for a given protein sequence, this paper presents a novel structure learning method that can predict a set of diverse contact maps for a given protein sequence, in which the best solution usually has much better accuracy than the first one. Our experimental tests show that for many test proteins, the best out of 5 solutions generated by our method has accuracy at least 0.1 better than the first one when the top L/5 or L/10 (L is the sequence length) predicted long-range contacts are evaluated, especially for protein families with a small number of sequence homologs. Our best solutions also have better quality than those generated by the two popular EC methods Evfold and PSICOV.
\end{abstract}
\IEEEpeerreviewmaketitle

\section{Introduction}
Protein contacts contain important information for protein folding and recent works indicate that one correct long-range contact for every 12 residues in the protein allows accurate topology level modeling \cite{kim2013one}. Thanks to high-throughput sequencing and better statistical assessment techniques, evolutionary coupling (EC) analysis for contact prediction has made good progress, which makes de novo structure prediction of some large proteins possible \cite{marks2011protein,nugent2012accurate,sulkowska2012genomics}. Nevertheless, contact prediction accuracy for many proteins is still low even if only the top $L$/10 ($L$ is the sequence length) predicted contacts are evaluated. \par

A number of contact prediction methods have been developed including \cite{burger2010disentangling}, PSICOV\cite{di2011there}, Evfold\cite{marks2011protein}, 
SVMSEQ\cite{wu2008comprehensive}, NNcon\cite{tegge2009nncon}, 
SVMcon\cite{cheng2007improved}, CMAPpro\cite{di2012deep},
PhyCMAP\cite{wang2013predicting} and Astro-Fold \cite{klepeis2003astro}. 
Some of them are pure EC-based unsupervised methods while others are supervised machine learning methods that integrate both EC and a variety of non-EC information. This paper focuses on the pure EC-based unsupervised approach, exploring one more way to improve residue evolutionary coupling (EC) analysis. EC analysis detects co-evolved residues from the MSA (multiple sequence alignment) of a protein family and then predicts two co-evolved residues to form a contact. This approach is based upon an observation that a pair of co-evolved residues is often found to be spatially close in the 3D structure. Along with many more sequences generated by high-throughput sequencing, some sophisticated global statistical methods (or structure learning methods), such as maximum entropy and probabilistic graphical models, are developed to infer residue co-evolution from MSA\cite{balakrishnan2011learning, cocco2013principal,lapedes2012using, marks2011protein, weigt2009identification}. These global statistical methods can differentiate direct from indirect residue couplings and thus, are more accurate than the traditional mutual information, which is a local statistical method, in predicting contacts.  Global statistical methods differ from local methods in that the former considers their relationship with all the other residues in determining if two residues are co-evolved or not while the latter ignores this kind of relationship. 
Representative tools of recent global statistical methods for contact prediction  include   Evfold\cite{marks2011protein}, PSICOV\cite{jones2012psicov} and GREMLIN\cite{balakrishnan2011learning}. See \cite{de2013emerging} for an excellent review of EC analysis.\par

Although quite a few new contact prediction methods have been developed recently,
existing methods predict only a single contact map (also called contact graph) for a protein sequence. 
Such a single solution may have low accuracy especially when the protein under prediction does not have a large number of sequence homologs, i.e., no sufficient information for the determination of the ground-truth contact map. This paper tackle the challenging contact prediction problem from another perspective. Analogous to ab initio folding that usually predicts a few possible 3D models for a given protein sequence, this paper presents a novel structure learning method that can predict $M$ (a small constant) diverse contact maps for a given protein sequence such that the best of the predicted contact maps usually has much better accuracy than the first one. In addition to ab initio folding, the idea of generating multiple solutions has been studied for other bioinformatics problems such as sequence and structure alignment and genome assembly, but not for contact prediction. All existing contact prediction methods generate only a single contact map for a protein sequence, although they may assign a confidence score to each predicted contact. \par

Our experimental tests show that for many test proteins, we only need to generate five different contact maps for a single protein in order to yield a contact map with much better accuracy than the first one and those generated by two popular EC analysis methods PSICOV and Evfold. This is especially true when the protein family under prediction has a small number of sequence homologs. This paper focuses on how to generate diverse $M$-best contact maps, but does not study how to re-rank them. A possible way to do so is that we may conduct protein folding simulation using each predicted contact map and then select the one resulting in the best protein 3D models according to some atom-level energy functions or model quality assessment methods.  \par

%{\color{red}maybe we need a little bit more introduction to make the first page is only about introduction}

\section{Methods}
\subsection{ Background }
In this section we introduce some notations and PSICOV\cite{jones2012psicov}, one state-of-the-art co-evolution analysis method for contact prediction. PSICOV formulates the contact prediction problem using Gaussian Graphical model (GGM). Our proposed method will also be based upon GGM, but the idea we propose here can be easily adapted to other models, such as the Potts model \cite{ekeberg2013improved} and the pseudo-likelihood model \cite{seemayer2014ccmpred}. \par

We run the buildali.pl program in the HHpred package \cite{soding2005hhpred} to find its sequence homologs and then build an MSA of these homologs.
We denote the MSA as $X = (x_1, x_2, ..., x_N)$, where $N$ is the number of sequence homologs and each sequence $x_i=(x_{i1}, x_{i2}, ..., x_{iL})$ is a vector with length $L$. Each $x_{ij}$ takes a value from a set $\Phi$ with cardinality 21 consisting of 20 amino acids and one gap symbol. So, $x_{ij}$ is a vector of 21 binary variables and each element in $x_{ij}$ indicates the presence or absence of a specific amino acid at row $i$ and column $j$. \par

Let $f_i(a)$ denote the frequency of amino acid $a$ at position $i$ and $f_{ij}(a,b)$ the frequency of one pair of amino acids $a$ and $b$ at positions $i$ and $j$. We may calculate  the $21L \times 21L$ covariance matrix for the MSA as follows.
\begin{align}
\label{p2}
	S_{ij}^{ab} = f_{ij}(ab) - f_i(a) f_j(b),
\end{align}
where the value of $S_{ij}^{ab}$ implies the correlation of amino acids $a$ and $b$ at positions $i$ and $j$. In this paper we also use $S_{ij}$ to denote the $21\times 21$ sub-matrix corresponding to two positions (columns) $i$ and $j$, and the value of the submatrix at position $(a, b)$ is $S_{ij}^{ab}$. \par

Given the empirical (or data) covariance matrix $S$ above, we can compute its inverse by minimizing the negative log-likelihood with a $l_1$ penalty term:
\begin{align}
\label{glasso}
	\Omega =& \arg \min_W F(W)  + \lambda \| W \| \nonumber \\
=& \arg \min_W tr(SW)-\ln \det(W) + \lambda \sum_{ij} |W_{ij}|
\end{align}
where $F=tr(SW)-\ln \det(W)$ is the negative log-likelihood of the observed sequences in an MSA, $\Omega$ is the estimated precision matrix (i.e., inverse of S), and $\lambda$ is a hyper-parameter controlling the sparsity level of $\Omega$. The structure of the resultant precision matrix $\Omega$ encodes the residue interaction pattern and can be used to derive a contact map \cite{jones2012psicov}. To efficiently optimize Eq.(\ref{glasso}), please refer to \cite{friedman2008sparse}. \par

PSICOV formulates the contact prediction problem using Eq. (\ref{glasso}) and predicts contacts from the estimated precision matrix using the $l_1$ norm of the submatrix corresponding to two positions (or columns) $i$ and $j$ (excluding the contribution of gaps), i.e.
\begin{align}
	\label{l1}
	\tilde{\Omega}_{ij} = \sum_{a,b \in \Phi} | \Omega_{ij}^{ab}|,
\end{align}
where $a$ and $b$ range over the 20 amino acids. That is, $\tilde{\Omega}_{ij}$ is a measure of interaction strength between positions $i$ and $j$. The higher $\tilde{\Omega}_{ij}$, the more likely there is a contact between $i$ and $j$. PSICOV also reduces the entropic and phylogenetic bias by correcting $\tilde{\Omega}_{ij}$ with average product correction (APC)\cite{dunn2008mutual}. Therefore the final prediction score can be represented as
\begin{align}
\label{post2}
	\tilde{\Omega}_{ij}^{corrected} = \tilde{\Omega}_{ij} - \frac{ \tilde{\Omega}_{i\cdot}\tilde{\Omega}_{\cdot j} }{ \tilde{\Omega}_{\cdot\cdot} }
\end{align}
where $\tilde{\Omega}_{i\cdot}$ is the mean of all the $\tilde{\Omega}_{ij}$ sharing position $i$, $\tilde{\Omega}_{\cdot j}$ is the mean of all the $\tilde{\Omega}_{ij}$ sharing position $j$  and $\tilde{\Omega}_{\cdot\cdot}$ is the mean of all the $\tilde{\Omega}_{ij}$. 
Finally, those residue pairs with large $\tilde{\Omega}_{ij}^{corrected}$ are predicted to form a contact.
\subsection{A new model for predicting diverse M-best contact maps}
For many proteins, a single solution may deviate a lot from the ground truth \cite{batra2012diverse}. 
Here we propose a method that can generate a set of alternative contact maps for a protein under prediction. We require that the set of solutions to be generated are diverse, so that the best solution may have much better accuracy. Note that since the search space is huge, it is still challenging to yield a solution with better accuracy even if we predict a small number (e.g., 5 or 10) of diverse solutions. \par

We generate a set of diverse solutions using an iterative process. Let $\Omega^1$ denote the first solution to Eq.(\ref{glasso}), i.e., the precision matrix generated by the popular method PSICOV. Then we generate a new solution $\Omega^2$ based on $\Omega^1$, such that the distance between them, $d(\Omega^1, \Omega)$, is larger than a desired value $\epsilon$, which quantifies the diversity of the alternative solutions. Next we generate $\Omega^3$ based on both $\Omega^1$ and $\Omega^2$. Generally speaking, in step $m+1$ we generate $\Omega^{m+1}$ based on all previous precision matrices $\Omega^1, \Omega^2, \dots, \Omega^m$, such that the new solution is different from all of them. More specifically, we will generate a new solution by solving the following optimization problem:
\begin{align}
\label{original_opt}
\min_\Omega & \ F(\Omega) + \lambda \sum_{ij} | \Omega_{ij}| \\
s.t. \ \  &d(\Omega^k, \Omega) \ge \epsilon, k=1, \dots, m,\nonumber
\end{align}
where $d(\Omega^k, \Omega)$ measures the difference between two alternative solutions $\Omega^k$ and $\Omega$. The new solution resulting from this formulation minimizes the negative log-likelihood of the observed sequences in the MSA, subject to the constraint that it must be different from previous solutions by a given cutoff $\epsilon$. Note that even though the previous results $\Omega_1, \Omega_2, \dots, \Omega_m$ are not explicitly expressed in the objective function, they impact the new solution $\Omega$ through the constraints. \par

We use the popular Hamming distance to measure the difference between two solutions, which is defined as follows.
\begin{align}
d(\Omega^k, \Omega) = (L^2  - \langle \delta_0(\Omega^k) ,  \delta_0(\Omega) \rangle )/2
\end{align}
where $L^2$ is the number of residue pairs (or submatrices) in $\Omega$, and $\delta_0$ maps a real-valued $21L \times 21L$ matrix  to a binary $L \times L$ matrix. More precisely, if the submatrix $\Omega_{ij}$ for residue pair $(i,j)$ is not 0, then  $\delta_0({\Omega_{ij}})$ takes value 1; otherwise -1. The operator $\langle\cdot, \cdot\rangle$ calculates the inner product of two matrices. The Hamming distance is always larger than or equal to zero. The smaller the distance is, the more similar contact maps implied by the two solutions.\par

However, it is challenging to optimize Eq.(\ref{original_opt}) with respect to this distance constraint function since $\delta_0(\Omega)$ is neither convex nor differentiable. To tackle this problem, we relax the Hamming distance as follows.
\begin{align}
\hat{d}(\Omega^k, \Omega) = - \langle \delta_0(\Omega^k), |\Omega|\rangle = - \sum_{ij} \delta_0(\Omega^k)_{ij} |\Omega_{ij}|,
\end{align}
Since at this point $\Omega^k$ is a constant, the above relaxation is actually a negative re-weighted $l_1$ norm of the precision matrix $\Omega$. Such a relaxation has the following desired properties:
\begin{itemize}
\item If there is a contact between $i$ and $j$ in the previous solution $\Omega^k$, the chance of there being a contact between $i$ and $j$ in $\Omega$ will be decreased.
\item If there is no contact between $i$ and $j$ in all previous $\Omega^k$, the chance of there being a contact between $i$ and $j$ in $\Omega$ will be increased.
\item The optimization is also easier since the relaxation is a convex function.
\end{itemize}

The first two properties ensure that the new solution will be different from all previous ones. The third property will not only simplify the optimization, but also significantly reduce the computational cost by using a specific procedure to be described in the next section. Finally, our original formulation (\ref{original_opt}) becomes
\begin{align}
\label{opt}
\min_\Omega & \ F(\Omega) + \lambda \sum_{ij} | \Omega_{ij}| \\
s.t. \ \  &\hat{d}(\Omega^1, \Omega) \ge \epsilon, k=1, \dots, m\nonumber
\end{align}
Where the continuous distance function $\hat d$ replaces the 0-1 Hamming distance $d$ in (\ref{original_opt}). 

\subsection{Optimization}
It is easy to prove that for every $\epsilon$ (details ignored due to space limit), there exists some $\mu(\epsilon)$ such that the problem (\ref{opt}) is equivalent to minimizing
\begin{align}
\label{new_opt}
G(\Omega) =  & F(\Omega)  + \lambda \| \Omega \|_1 + \sum_{k=1}^m -  \mu_i(\epsilon) \hat{d}(\Omega^m, \Omega)  \\
 =  &tr(S\Omega) - \ln \det (\Omega) + \sum_{ij} (\lambda - \sum_{k=1}^m \mu_k(\epsilon) \delta_0(\Omega^k)_{ij} ) |\Omega_{ij}| \nonumber
\end{align}

Compared to (\ref{opt}), the new formulation (\ref{new_opt}) is appealing since it has no constraints and also has the same form as the original formulation (\ref{glasso}), which has been extensively studied and many solvers are developed. This allows us to apply existing solvers to solve our new formulation (\ref{new_opt}). The hyper parameters $\mu_k(\epsilon)$ controls how far away we want our solution to deviate from the previous solution $\Omega^k$. To prevent overfitting, we set all $\mu_k$ to be the same. \par

Note that all the $\mu_k$ shall be upper bounded by $\lambda/m$. Otherwise, $\lambda'_{ij} = \lambda - \sum_{k=1}^m \mu \delta_0(\Omega^k)_{ij}$ becomes negative and thus, formulation(\ref{new_opt}) becomes non-convex. 
The intuition underlying this upper bound is that the difference between two solutions shall not be too big, which makes sense since all the solutions shall minimize the negative log-likelihood of the observed sequences in an MSA. 

Finally, we generate a new contact map by minimizing the following function.
 \begin{align}
\label{final_opt}
G(\Omega)  =& tr(S\Omega) - \ln \det (\Omega) + \sum_{ij} (\lambda - \sum_{k=1}^m \mu \delta_0(\Omega^k)_{ij} ) |\Omega_{ij}| \nonumber \\
=& tr(S\Omega) - \ln \det (\Omega) + \sum_{ij} \lambda'_{ij} |\Omega_{ij}| \\
&s.t. \ \  0 \le \mu < \lambda/m  \nonumber
\end{align}
Empirically we can solve (\ref{final_opt}) using graphical Lasso \cite{friedman2008sparse} with time complexity of $O(p^3)$, where $p=21L$ is dimension of $\Omega$. Nevertheless, if we want to generate $m$ alternative solutions, the running time will be $m$ times slower than PSICOV, i.e., generating a single solution.\par

{\bf Speedup.} Here we describe a novel procedure to significantly reduce the computation time, by exploiting the facts that $\Omega$ is very sparse and there is still some similarity between two alternative solutions (although we want them to be diverse). 

For a consecutive pair $k$ and $k+1$, we can assume that the difference between $\lambda^k$ and $\lambda^{k+1}$ is not too large because of the update function for $\lambda_{ij}'$ in (\ref{final_opt}). Therefore, we can estimate the new precision matrix $\Omega^{k+1}$ by first copying from $\Omega^k$ and then updating a small number of entries. Based on this observation, we propose following procedure. \par

In the first step, we apply the Quadratic Inverse Covariance (QUIC) method \cite{hsieh2011sparse} to optimize (\ref{final_opt}) to generate the initial solution. Unlike the optimization methods used in PSICOV, QUIC has a super linear convergence rate because it uses the second order information. At each iteration, QUIC approximates $F(\Omega)$ by its second-order Taylor expansion, computes the Newton direction, and descends at one coordinate. \par

In the second step, which is key for speedup, we identify which variables need to be updated. It is not hard to prove that if the following condition holds in current iteration,  we do not need to update the submatrix $\Omega_{ij}$ in the remaining iterations.
\begin{align}
\label{id_var}
|\nabla_{ij} F(\Omega)| \le \lambda_{ij} \text{ \ and \ } \Omega_{ij} = 0
\end{align}
%therefore we only need to update a small amount of variables rather than all of the $p^2$ variables.
Where $\nabla_{ij} F(\Omega)$ is the gradient of $F(\Omega)$ at position $(i,j)$. Let $S_{fixed}$ denote the set of variables which do not need update and $S_{free}$ the complement of $S_{fixed}$. 
Due to the similarity between two consecutive alternative solutions, we would expect that the number of variables in $S_{free}$ is small. 
In practice we only need to update less than one percent of matrix entries at each step, which significantly improves the running time compared to simply running QUIC or graphical Lasso multiple times. That is, our method can quickly generate the alternative solutions after computing $\Omega^1$ by the update rule in (\ref{id_var}). We summarize our procedure in algorithm 1. \par

\begin{algorithm}[ht]
\caption{Algorithm for diverse M-Best contact maps}
\label{alg2}
\begin{algorithmic}[1]
\State \text{Input \ } $\lambda$, $\mu$, $m$
\For {$k=1, ..., m$}
	\If {k =1}
		\State $\Omega^1$ = QUIC($S$, $\lambda$)
	\Else
		\While { not converge }
			\State Compute $\lambda_{ij} = \lambda - \sum_{r=1}^{k-1} \mu \delta_0(\Omega^r)_{ij}  \ \forall \ i, j$
		    	\State Identify $S_{fixed}$ and $S_{free}$ by (\ref{id_var})
			\State Compute the Newton direction $\Delta$ on $S_{free}$
			\State Compute step size $\alpha$ by linear search
			\State Update $\Omega^k = \Omega^k - \alpha \Delta$ %{\color{red} u did not say how to calculate $\alpha$}
		\EndWhile
	\EndIf
\EndFor
\State{\textbf{output} $\Omega^1, \Omega^2, ..., \Omega^m$}
\end{algorithmic}
\end{algorithm}

To compute the update direction and step size in lines 9 and 10 of our algorithm, we need to maintain a temporary matrix to compute the Hessian matrix of $F(\Omega)$ efficiently. 
See \cite{hsieh2011sparse} for more technical details. The algorithm runs very fast after computing $\Omega^1$ due to step 8.

\subsection{Contact Selection by Nuclear Norm}
\begin{figure}[h]
\centerline
{\includegraphics[width=80mm]{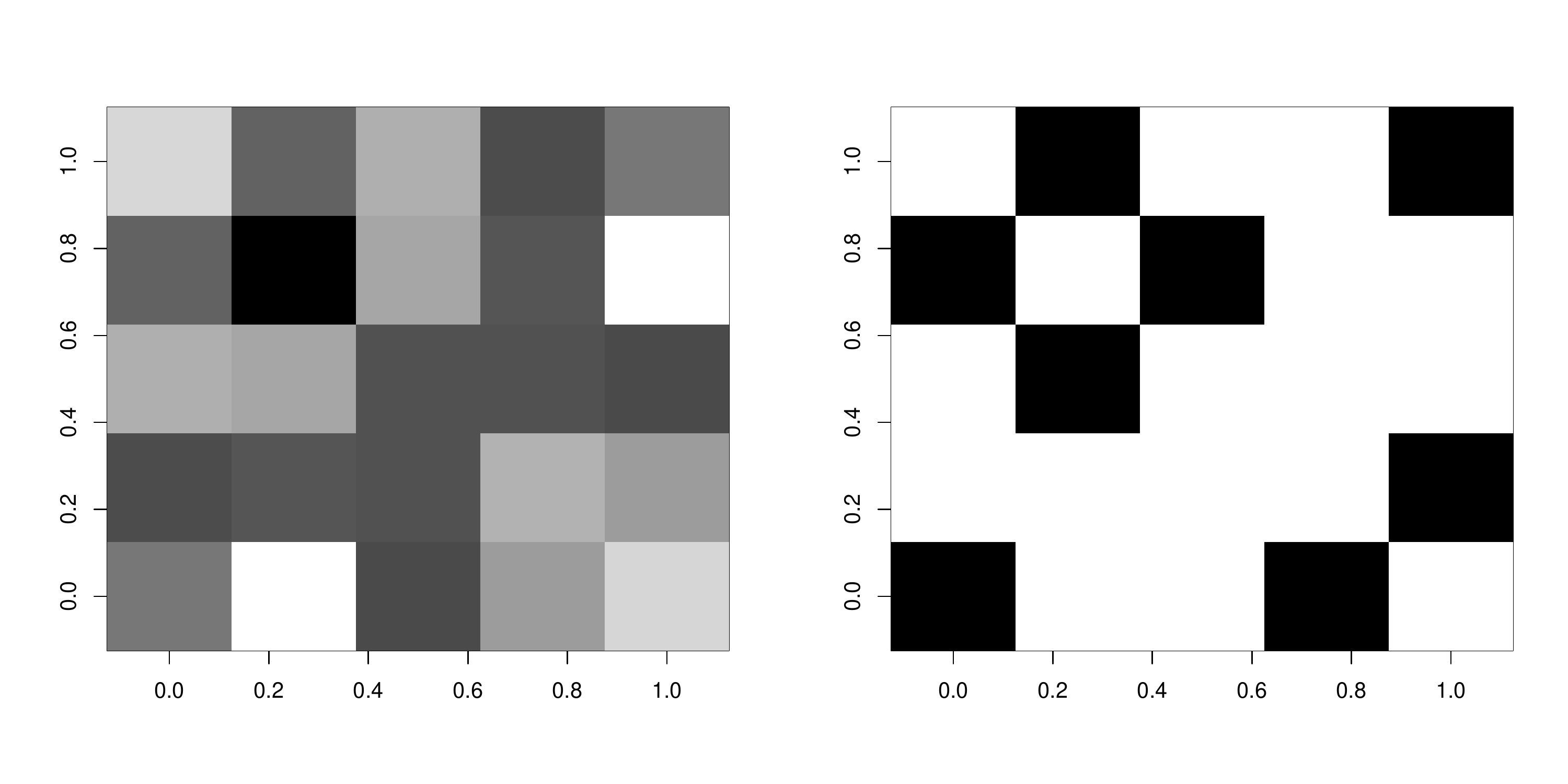}}
\caption{The figure shows two artificial matrices with different patterns, but same $l_1$ norm. The right one has a much stronger interaction pattern than the left one. }\label{NuclearNorm}
\end{figure}

After computing the precision matrix, existing methods such as PSICOV select top predicted contacts based on the $l_1$ norm of each sub-matrix $\Omega_{ij}$(see Eq.\ref{l1}), which can be interpreted as the interaction strength between two positions (or columns). Here we propose a new method that uses the nuclear norm of each sub-matrix $\Omega_{ij}$ to measure the interaction strength of two positions. That is,
\begin{align}
\label{nuclear}
\tilde{\Omega}_{ij} = \sum_{i=1}^{20} \sigma_i
\end{align}
where $\sigma_i$ is the $i$-th singular value of matrix $\Omega_{ij}$. Again, we do not use the row and column in $\Omega_{ij}$ corresponding to the gap in calculating the nuclear norm. The nuclear norm is better than $l_1$ norm in capturing the sparsity pattern in each sub-matrix. 
Figure \ref{NuclearNorm} shows two artificial $5 \times 5$ matrices with entry values ranging from 0 to 1 shown by different colors. The darker the color, the higher the value. Both matrices have the same $l_1$ norm. However, it is easy to see that the right matrix has a much stronger interaction pattern than the left one, which is consistent with their nuclear norms. The nuclear norms of the left and right matrices are 8.36 and 18.16, respectively. On average the nuclear norm is better than  $l_1$ norm in measuring residue interaction strength since the former can better differentiate the sparsity pattern (right) from noise (left).

\section{Results}
We evaluate our method by examining the improvement of the best solution over the first solution. We will also show that the best solutions generated by our method are also better than PSICOV and Evfold. 
We used three test sets: (1) The CASP10 test targets; (2) the PSICOV benchmark (150 Pfam families) and a subset of PDB25, which contain proteins with a large number of sequence homologs; and (3) a subset of PDB25 consisting of proteins with a small number of sequence homologs and thus, is very challenging. The PSICOV benchmark already contains multiple sequence alignment (MSA) from the Pfam database. For other test proteins, their multiple sequence alignments are generated using buildali.pl in the HHpred package \cite{soding2005hhpred}, which runs PSI-BLAST \cite{altschul1997gapped} to find sequence homologs.

We consider only medium- and long-range contacts since they are much more important than short-range contacts. Following PSICOV and other methods, we say there is a contact between two residues if their $C_\beta$ Euclidean distance is less than 8$\AA$. A contact of two residues at sequence positions $i$ and $j$ is short-, medium- and long-range if $|i-j|$ is less than 12, between 12 and 24 or larger than 24, respectively.

Following PSICOV, we set the sparsity control parameter  $\lambda$ to 0.01. The diversity parameter $\mu$ is set to 0.0015 for all test cases. In this paper, for each protein under prediction, we predict only 5 different contact maps although it may lead to better accuracy by producing more solutions.

To ensure fair comparison, we employ the same pre- and post-processing procedures used by PSICOV, except that we use nuclear norm to measure the interaction strength rather than $l_1$ norm. Briefly, to reduce the impact of redundant sequences, we apply the same sequence weighting method as PSICOV. In particular, duplicate sequences are removed and columns containing more than 90\% of gaps are deleted. The sequence is weighted using a threshold of 62\% sequence identity. Similar to PSICOV and plmDCA\cite{ekeberg2013improved}, average-product correction (APC) \cite{dunn2008mutual} is applied to correct bias (see Eq. \ref{post2}).

\subsection{Performance on 123 CASP10 targets}
\subsubsection{Advantage of nuclear norm for contact selection}
We first show that nuclear norm is better in selecting top contacts. We run PSICOV to generate precision matrices and then employ nuclear norm and $l_1$ norm to select top $L/5$ contacts ($L$ is sequence or MSA length), respectively. Both norms have similar performance in selecting medium-range contacts. The $l_1$ norm and the nuclear norm have accuracy 0.256 and 0.255, respectively. For long-range contacts, the nuclear norm and $l_1$ norm have accuracy 0.237 and 0.225, respectively.

\subsubsection{The best solution vs. the first solution}
\begin{figure}[t]
\centerline
{\includegraphics[width=60mm]{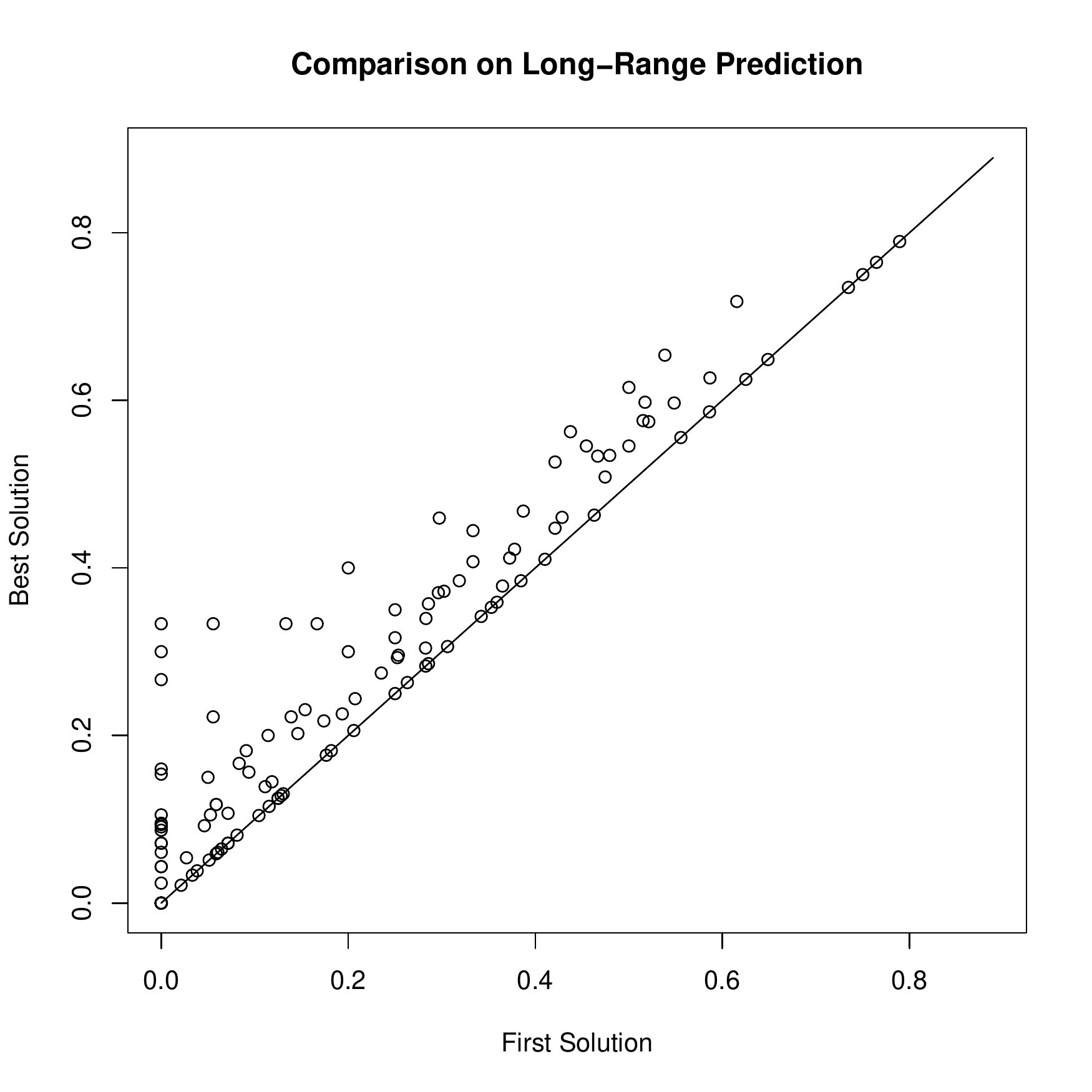}}
\caption{The figure illustrates the one-to-one comparison on top $L/5$ long-range contact prediction for CASP10 proteins. The X-axis is the accuracy for first solution and the Y-axis is that of the best solution. }\label{CASP10}
\end{figure}

When top $L/5$ predicted contacts are evaluated, the average medium-range prediction accuracy of the best (out of 5) solutions is 0.285, which is 11.51\%  better than the first solutions (0.255). The best solutions have long-range prediction accuracy 0.279, which is a 17.7\% better than the first solutions (0.237). See Figure \ref{CASP10} for a detailed one to one comparison. When only the top $L/10$ predictions are evaluated, the first solutions have average medium- and long-range accuracy of 0.301 and 0.275, respectively. By contrast, the best solutions have average accuracy of 0.335 and 0.322, respectively. Overall, the best solutions tend to have a larger advantage over the first solutions in predicting long-range contacts, which is desirable since long-range contacts are the most useful for protein folding. In particular, for 32 out of the 123 test proteins, their best solutions have long-range accuracy at least 0.1 better than the first solutions. For 6 test proteins, their best solutions have long-range accuracy at least 0.2 better than the first ones. Note that the average accuracy of this test set is not very high mainly because the CASP10 set contains many test proteins with a very small number of sequence homologs.

%\begin{figure}[t]%[!tpb]%figure1

\subsubsection{Performance with respect to the number of solutions}
\begin{figure}[t]
\centerline
{\includegraphics[width=60mm]{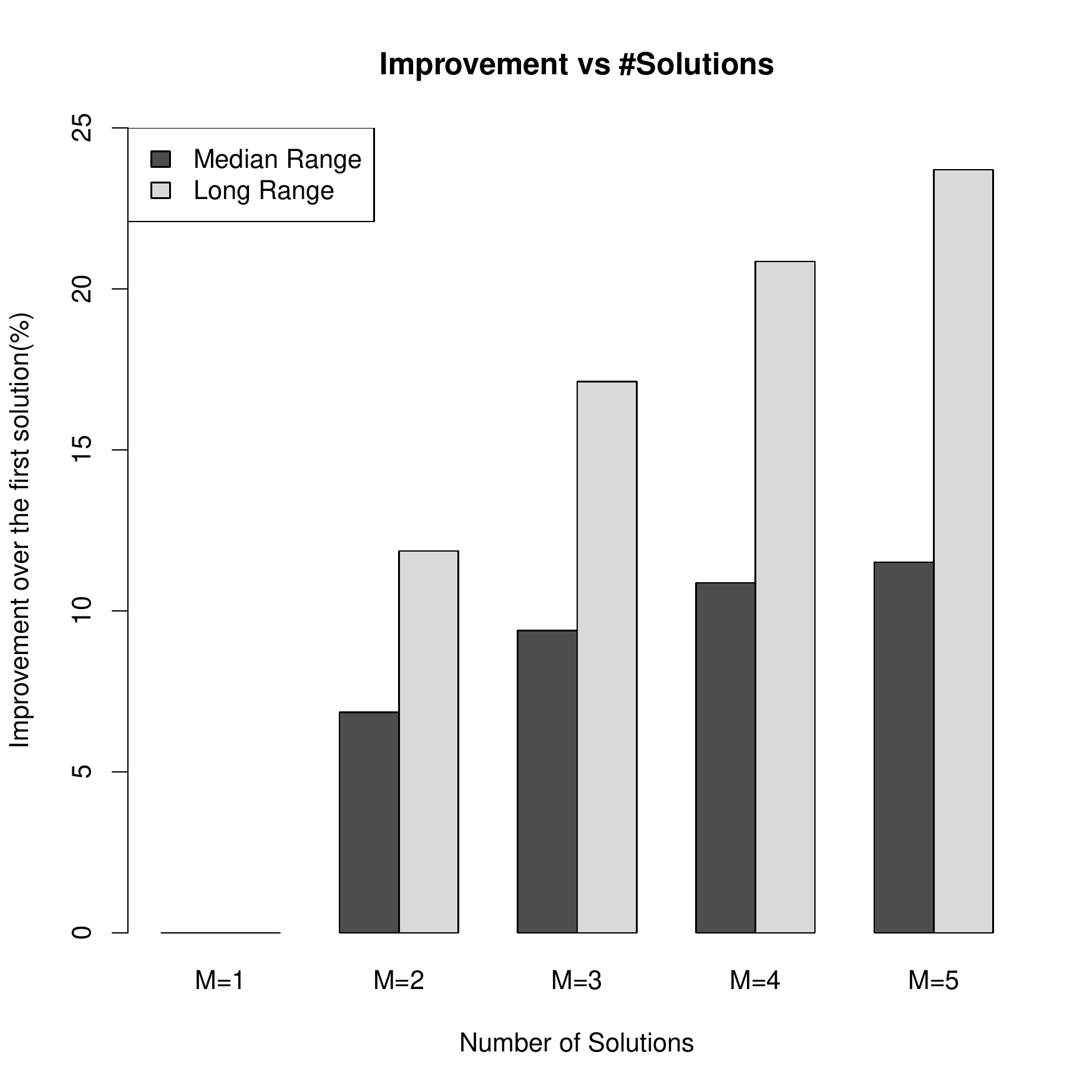}}
\caption{The accuracy improvement of the best solutions with respect to the number of solutions generated for a single protein. This figure shows that long-range prediction accuracy increases significantly with respect to the number of solutions.}\label{PerfWRTNumberOfSolutions}
\end{figure}

As shown in Figure \ref{PerfWRTNumberOfSolutions}, the more solutions we generate, the better accuracy can be obtained by the best solutions. The long-range prediction accuracy can be improved by 24\% when only 5 diverse solutions are generated. Even if only two solutions are generated for a protein, we can obtain 6.85\% and 11.86\% improvement, respectively, in medium- and long-range contact prediction.
%\begin{figure}[t] %[!tpb]%figure2

\subsubsection{Performance with respect to the amount of homologous information} 
Let $M_{eff}$ denote the number of non-redundant sequence homologs available for a protein (family) under prediction.
Following \cite{morcos2011direct}, $M_{eff}$ is calculated by
\begin{equation}
\label{meff}
M_{eff} = \sum_i \frac{1}{\sum_j M_{i,j}}
\end{equation}
where $i$ and $j$ represent two sequence homologs of the protein (family) under prediction and $M_{ij}$ is a binary value indicating whether $i$ and $j$ are redundant. In particular, $M_{ij}=1$ if and only if $i$ and $j$ share more than 70\% of sequence identity; otherwise 0. $M_{eff}$ measures the amount of homologous information available for a protein (family), so it can be used to evaluate the difficulty level of a test protein (family). 

\begin{figure}[!h]
\centerline
{\includegraphics[width=100mm]{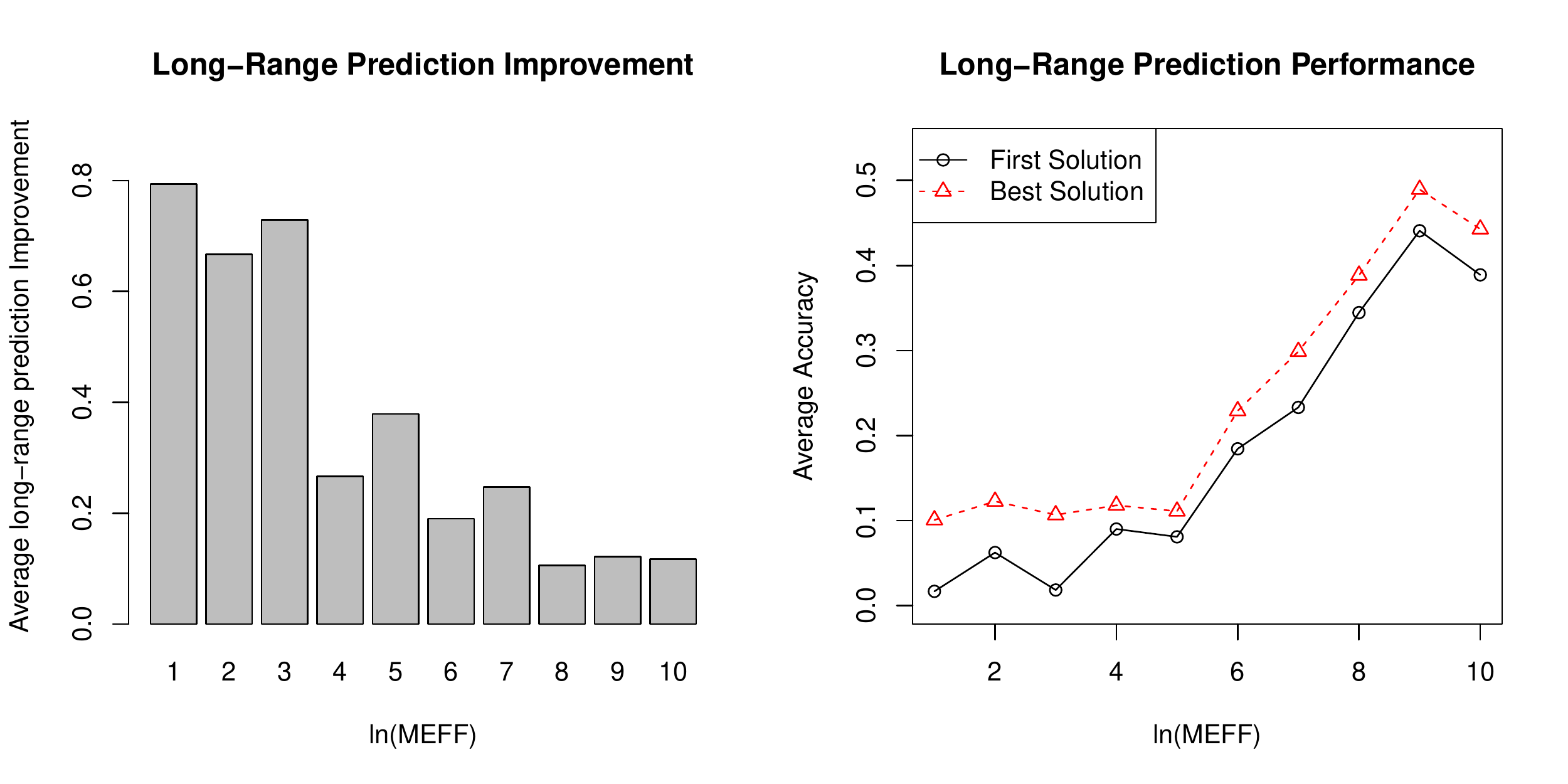}}
\caption {The left picture shows the relative improvement in long-range prediction accuracy of the best solutions over the first ones with respect to $\ln M_{eff}$, tested on the CASP10 set. The right picture shows the long-range accuracy of the best and first solutions.}\label{caspmeff}
\end{figure}

We divide the 123 CASP10 test proteins into 10 groups by their $\ln M_{eff}$ values: $1\sim 2, \dots, 9\sim 10, \geq10$, and then compute the average improvement in each group. As shown in Figure \ref{caspmeff}, the best out of 5 solutions have much larger advantage over the first solutions for proteins with smaller $M_{eff}$ values. The reason is that with a limited amount of homologous information, the first predicted contact map may differ a lot from the ground truth.

%\begin{figure}[H]
%\centerline
%{\includegraphics[width=80mm]{caspcmp.pdf}}
%\caption{The figure shows the improvement of our M-Best model on sequences with different $\log M_{eff}$. on both medium range contact prediction (left) and long range contact prediction (right) on CASP 10 data set.. }\label{meff}
%\end{figure}

\subsection{Performance on proteins with a large number of sequence homologs}
We build an “easy” data set, which consists of 149 test proteins in the PSICOV benchmark with $\ln M_{eff}$ between 7 and 10 and 140 proteins in PDB25 with $\ln M_{eff }$ between 5 and 7. The PDB25 list is downloaded from PISCES\cite{wang2003pisces}. Any two proteins in PDB25 share less than 25\% sequence identity. Compared to the CASP10 set, the test proteins in this data set on average have many more sequence homologs, so their contacts are relatively easier to predict.

\begin{figure}[!h]
\centerline
{\includegraphics[width=60mm]{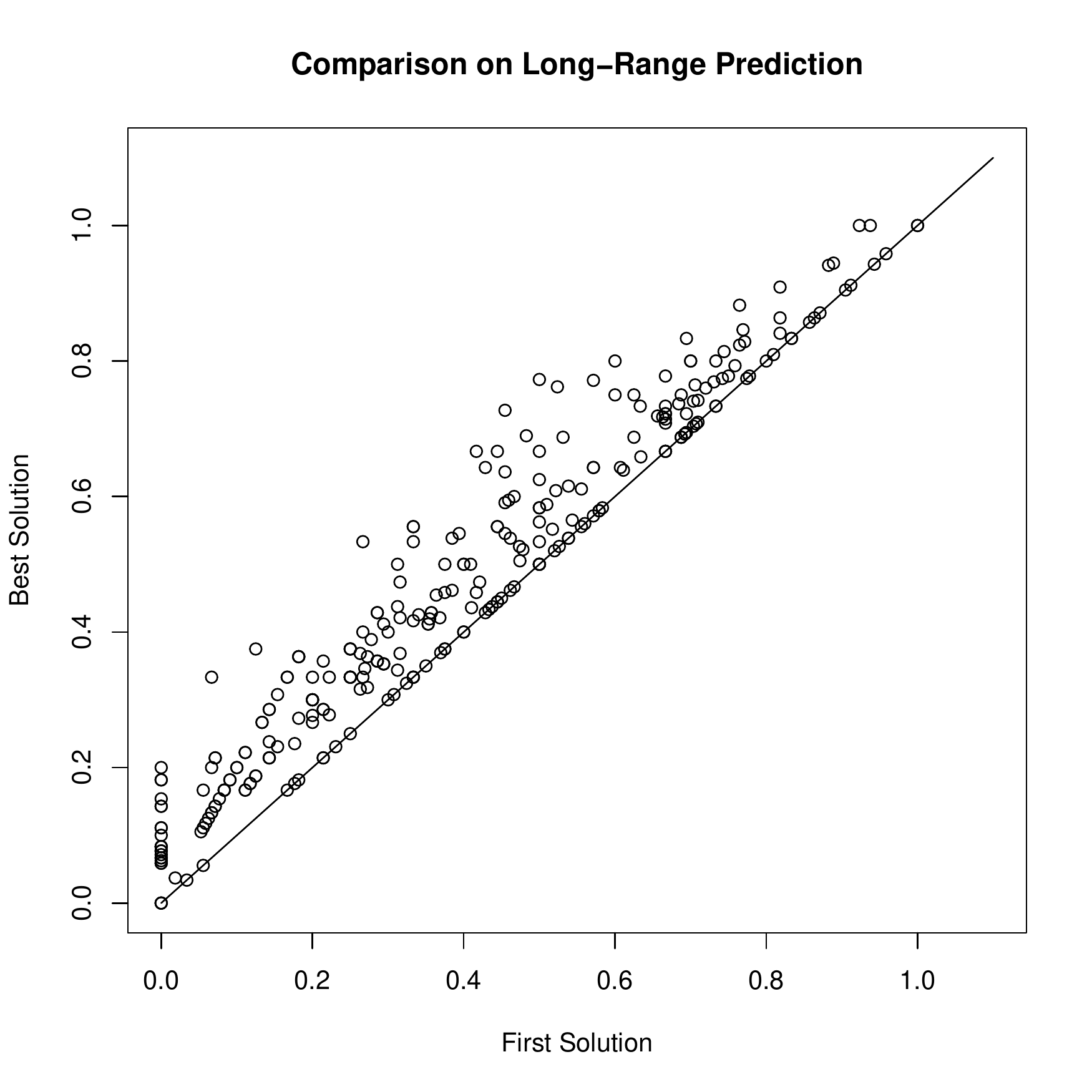}}
\caption {The above figure shows one-to-one comparison of the best and first solutions in terms of top $L/5$ long-range contact prediction accuracy for proteins with $\ln M_{eff}$ between 5 and 10.}\label{easy}
\end{figure}

Overall, when top $L/5$ predictions are evaluated, the best solutions are better than the first solutions by around 20\% for both medium- and long-range prediction. More specifically, for medium-range predictions, the best and first solutions have accuracy 0.381 and 0.315, respectively. For long-range prediction (see Figure \ref{easy}), the best and first solutions have accuracy 0.464 and 0.391, respectively. In particular, for 111 out of 289 test proteins, the best solutions have accuracy at least 0.1 better than the first solutions for long-range prediction. For 23 test proteins, the best solutions have accuracy at least 0.2 better than the first ones for long-range prediction.

When top $L/10$ predictions are evaluated, the average accuracy of the first solutions are 0.380 and 0.419, respectively, for medium- and long-range contacts, while the best solutions have accuracy 0.445 and 0.499, respectively. 

\subsection{Performance on proteins with a small number of sequence homologs}
\begin{figure}[t]
\centerline
{\includegraphics[width=60mm]{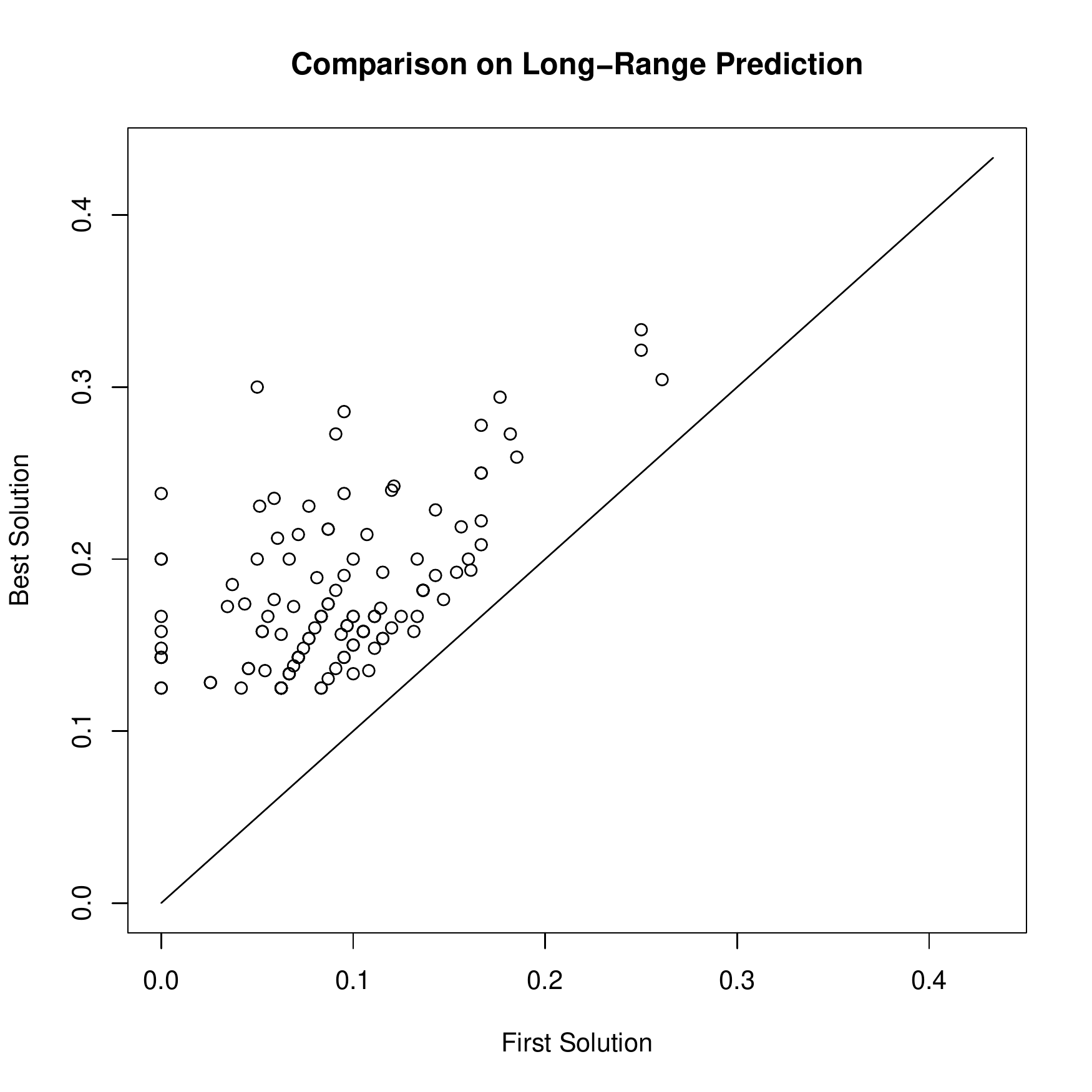}}
\caption {The above figure shows one-to-one comparison on top $L/10$ long-range contact prediction for proteins with $\ln M_{eff}$ less than 5.}\label{hard}
\end{figure}

We build a data set of 122 test proteins from PDB25, all of which have $\ln M_{eff}$  smaller than 5. This set is much more challenging to predict since the test proteins have only limited amount of homologous information. When top $L/10$ predicted contacts are evaluated, the first solutions has average accuracy of 0.111 and 0.133 on medium-range and long-range contacts, respectively, while the best solutions have accuracy of 0.172 and 0.214, respectively. As shown in Figure \ref{hard}, for 58 out of 122 test proteins, the best solutions have long-range accuracy at least 0.1 better than the first solutions. For 15 test proteins, the best solutions have long-range accuracy at least 0.2 better than the first ones. \par

When top $L/5$ predictions are evaluated, the average accuracy of the first solutions are 0.089 and 0.090, respectively, for medium- and long-range contacts, while the best solutions have accuracy 0.143 and 0.180, respectively. 

\subsection{Comparison with other co-evolution-based methods}

We compare the accuracy of our M-best method with a couple of popular co-evolution methods such as PSICOV and Evfold \cite{morcos2011direct}. PSICOV is based upon Gaussian Graphical Model (GGM) and Evfold is derived by maximum entropy. 
% and plmDCA uses a pseudo-likelihood method. Since plmDCA does not assume Gaussian distribution of the observed sequences, it generally outperforms PSICOV. 
We ran both PSICOV and Evfold by the default parameters suggested by their respective publications. 
Table I shows the average long-range prediction accuracy of PSICOV, Evofld and our method on the CASP10 set.
As shown in this table, our M-best method outperforms PSICOV and Evfold. 
%and has a similar performance as plmDCA. 
Our method is also based upon GGM, so its comparison with PSICOV implies that our method indeed can yield better accuracy by producing only a very small number of alternative solutions for an individual protein. 
%Comparison with plmDCA implies that GGM can yield similar performance as the pseudo-likelihood model as long as we generate 5 different solutions for the GGM.
In principle we may apply our M-best idea to the pseudo-likelihood model, which shall result in better accuracy than its corresponding tools such as plmDCA \cite{ekeberg2013improved} and CCMpred \cite{seemayer2014ccmpred} and further advance protein contact prediction. 

\begin{table}[t]
\caption{Contact prediction accuracy of PSICOV, Evfold and our method on the CASP10 set.}
\centering
\begin{tabular}{ccccc}
\toprule
               & PSICOV &  Evfold  & Our M-best method \\
\midrule
L/5 (long)     & 0.225 &   0.225   & 0.281 \\
\midrule
L/10 (long)    & 0.276 &   0.257   & 0.322 \\
\midrule
L/5 (median)  & 0.259 &   0.249    & 0.285 \\
\midrule
L/10 (median) & 0.310 &   0.294    & 0.335 \\
\bottomrule
\end{tabular}
\end{table}

\subsection{Case Study}
This section describes some specific cases on which the best solution is much better than the first one. One observation is that the best solutions tend to correctly predict more contacts in beta-sheet proteins. A specific example is PF07686 \cite{bateman2004pfam}, which belongs to Immunoglobulin V-set domain \cite{satow1986phosphocholine} and Ig-like domains resembling the antibody variable domain. V-set domains are found in diverse protein families, including immunoglobulin light and heavy chains. In addition, this domain belongs to SCOP family  b.1.1.1 \cite{andreeva2008data}, which is an all-beta protein in greek-key shape that forms a sandwich consisting 7 strands in 2 sheets. The best out of 5 solutions correctly  predicts 17 out of 22 long-range contacts, while the first solution only 11 of them.

Another example is PF01300, which belongs to Telomere recombination domain \cite{teplova2000structure} and has been shown to bind preferentially to dsRNA. It has also been shown to be required for telomere recombination in yeast. This domain belongs to SCOP family d.115.1.1, which is an alpha and beta protein (a+b) in YrdC-like shape containing two additional strands in the C-terminal extension. The first solution  correctly identifies 20 out of 35 long-range contacts, while our best solution identifies 27.

The CASP10 target T0688-D1 is a LRR (Leucine-rich repeat) domain \cite{kobe2001leucine} and a protein structural motif that forms an α/β horseshoe fold. It is composed of repeating 20-30 amino acid stretches that are unusually rich in the hydrophobic amino acid Leucine. In addition, this domain belongs to SCOP family c.10.2.1, which is capped at the N-end with a truncated EF-hand sub domain \cite{nakayama1994evolution}. The best out of 5 solutions can correctly identify 17 out of 37 native long-range contacts while the first solution only 11 contacts.

%%%%%%%%%%%%%%%%%%%%%%%%%%%%%%%%%%%%%%%%%%%%%%%%%%%%%%%%%%%%%%%%%%%%%%%%%%%%%%%%%%%%%
%
%     please remove the " % " symbol from \centerline{\includegraphics{fig01.eps}}
%     as it may ignore the figures.
%
%%%%%%%%%%%%%%%%%%%%%%%%%%%%%%%%%%%%%%%%%%%%%%%%%%%%%%%%%%%%%%%%%%%%%%%%%%%%%%%%%%%%%%

\section{Conclusion \& Discussion}
In this paper, we have presented a general structural learning method that can predict a set of diverse contact maps from an MSA. Compared to PSICOV, we only introduce one extra hyper-parameter to control the diversity of the predicted contact maps. Experiments confirm that our method can yield much better contact maps by generating only 5 different solutions,  especially when the protein under consideration has a small number of sequence homologs. Our method works particularly well for contacts in beta-sheet proteins.

Currently we just fix the value of the hyper-parameter for the diversity control. In the future we are going to choose such a hyper-parameter depending on the number of non-redundant sequence homologs in a protein family. That is, when a family contains fewer sequence homologs, we shall allow more diversity due to less information available in the multiple sequence alignment. Otherwise, we shall allow less diversity.

Although our method generates only 5 different solutions, the best solutions usually have better accuracy than the first ones. One possible reason may be the fact that PSICOV model is biased since it assumes Gaussian distribution of an MSA, which is not true when only a small number of sequence homologs are available. It was noted that the solutions to the single MAP or maximize-likelihood problem is often biased, so providing multiple diverse solutions can often help. Such phenomena occurred in a variety of computational biology problems, ranging from SNP data analysis \cite{silberstein2013system}, to sequence alignment\cite{do2006contralign}, to protein threading{\cite{ma2012conditional}, to protein structure alignment \cite{wang2013protein,zhang2005tm}, and to ab initio folding\cite{eswar2006comparative,rohl2004protein}. The strategy we present here could be directly used to some of the above-mentioned problems and provide multiple informative solutions.

Note that in this paper our intention is not to develop the best contact prediction method that outperforms all the existing methods. Instead, this paper studies how we can make the best use of co-evolutionary information and advance the Gaussian Graphical model for contact prediction. To achieve the best contact prediction accuracy, it usually needs to combine both co-evolutionary and non-co-evolutionary information. For example, MetaPSICOV \cite{jones2015metapsicov}, the best contact predictor in CASP11 (the 11th Critical Assessment of Structure Prediction), is a supervised learning method that integrates four pure co-evolution-based methods (including PSICOV and Evfold) and a variety of non-co-evolutionary information. Such a supervised learning method may not work well on membrane proteins since it is trained by globular proteins. By contrast, a pure co-evolution-based unsupervised method can work on any proteins. Similar to the spirit of MetaPSICOV, we may also develop a supervised learning method to integrate our M-best method, other co-evolution methods and some non-co-evolutionary information to further advance contact prediction.

\section*{Acknowledgment}
The authors are grateful to the financial support from National Institutes of Health R01GM0897532 and National Science Foundation CAREER award CCF-1149811



%\bibliographystyle{IEEEtran}
\bibliographystyle{plain}
\bibliography{document}

%\begin{thebibliography}{1}


%\end{thebibliography}

\end{document}